# A Matrix Formulation of Einstein's Vacuum Field Equations

Walter J. Wild
Department of Astronomy & Astrophysics
University of Chicago
5640 S. Ellis Ave.
Chicago, Illinois 60637
walter@babylon.uchicago.edu

**ABSTRACT**

We develop a correspondence between arbitrary tensors and matrices based on the use of Kronecker products and associated identities. Utilizing the rules of matrix differentiation we derive the vacuum Einstein field equations as a differential-matrix equation. This formulation may facilitate their efficient use in numerical relativistic models.

The general covariant field equations of general relativity find a natural representation using tensors. They consist of ten independent nonlinear second-order differential equations for the metric tensor components $g_{\mu\nu}$ in terms of the four space-time coordinates, and in a vacuum reduce to the vanishing of the Ricci tensor $R_{\mu\nu} = 0$, which is obtained from the only meaningful contraction of the Riemann curvature tensor $R^{\varepsilon}{}_{\mu\nu\sigma}$.[1] In the last several decades many researchers have worked with the field equations recast using the tools of differential forms[2] (the exterior calculus) and in the context of the null-tetrad, or Newman-Penrose, formalism.[3] Here we shall demonstrate that the vacuum field equations can be expressed as a system of first-order differential-matrix equations.

Tensors are convenient tools based on indicial structures and the summation convention, and as such enable N-dimensional arrays to be easily manipulated via transformation between upper (contravariant) and lower (covariant) indices and via contraction of indices. A matrix formulation regards all objects as two-dimensional structures which obey strict manipulative rules. However, equations posed in matrix form have a substantial repertoire of estimation and analyses tools available which can be utilized in numerical solutions; this may be especially important in numerical relativistic studies. As emphasized by E.T. Bell, matrices are "...a trivial notational device (which) may be the germ of a vast theory having innumerable applications."[4] Matrices may be thought of as very powerful organizational tools blessed with wide-ranging formal mathematical insights.

The connection between a $4k \times 4m$ matrix $\mathbf{A}$, for four space-time dimensions, and tensor $A^{\beta_1 \beta_2 \ldots \beta_m}{}_{\alpha_1 \alpha_2 \ldots \alpha_k}$, where for each $\alpha_j$ or $\beta_j$ the subscripted index $j \in \{1,2,3,4\}$, is established via the Kronecker product, "$\otimes$", defined between matrices and vectors.[5] The formalism is easily generalized to an arbitrary number of dimensions, j, so $\mathbf{A}$ is a $jk \times jm$ matrix. For any two matrices $\mathbf{A}$ and $\mathbf{B}$, the Kronecker product $\mathbf{A} \otimes \mathbf{B}$ has every element of $\mathbf{A}$ multiplied by the entire matrix $\mathbf{B}$, i.e.,

$$\mathbf{A} \otimes \mathbf{B} = \begin{bmatrix} a_{11}\mathbf{B} & a_{12}\mathbf{B} & a_{13}\mathbf{B} & \cdots & a_{1m}\mathbf{B} \\ a_{21}\mathbf{B} & a_{22}\mathbf{B} & a_{23}\mathbf{B} & \cdots & a_{2m}\mathbf{B} \\ a_{31}\mathbf{B} & a_{32}\mathbf{B} & a_{33}\mathbf{B} & \cdots & a_{3m}\mathbf{B} \\ & & \cdots & & \\ a_{n1}\mathbf{B} & a_{n2}\mathbf{B} & a_{n3}\mathbf{B} & \cdots & a_{nm}\mathbf{B} \end{bmatrix},$$

for an $n \times m$ matrix $\mathbf{A}$. This operation greatly increases the dimension of $\mathbf{A} \otimes \mathbf{B}$; in general the Kronecker product is not commutative so $\mathbf{A} \otimes \mathbf{B} \neq \mathbf{B} \otimes \mathbf{A}$. Consider a set of arbitrary four-element column vectors $\{\mathbf{a}_{\alpha_1}, \mathbf{a}_{\alpha_2}, \ldots, \mathbf{a}_{\alpha_k}; \mathbf{b}_{\beta_1}, \mathbf{b}_{\beta_1}, \ldots, \mathbf{b}_{\beta_m}\}$, where the subscripts designate different vectors corresponding to the index and not components, the dimensions of $\mathbf{A}$ and ordering of elements in $\mathbf{A}$ is specified by the Kronecker product ordering of these vectors in the scalar $\psi$, i.e.,

$$\psi = (\mathbf{a}^T_{\alpha_1} \otimes \mathbf{a}^T_{\alpha_2} \otimes \cdots \otimes \mathbf{a}^T_{\alpha_k}) \, \mathbf{A} \, (\mathbf{b}_{\beta_1} \otimes \mathbf{b}_{\beta_2} \otimes \cdots \otimes \mathbf{b}_{\beta_m}) \rightarrow \mathbf{A} = A^{\beta_1 \beta_2 \ldots \beta_m}{}_{\alpha_1 \alpha_2 \ldots \alpha_k} \quad . \tag{1}$$

We call this the $\psi$-representation of a matrix $\mathbf{A}$.

Interchanging any two indices in the tensor $A^{\beta_1\beta_2\cdots\beta_m}{}_{\alpha_1\alpha_2\ldots\alpha_k}$ is equivalent to interchanging the corresponding indexed vectors in $\psi$ and reordering the elements in $\mathbf{A}$ such that $\psi$ retains its same numerical value, i.e., remains invariant. Raising or lower indices in $A^{\beta_1\beta_2\cdots\beta_m}{}_{\alpha_1\alpha_2\ldots\alpha_k}$ corresponds to shifting the corresponding indexed four-vector to the proper location within the Kronecker products on each side of the matrix $\mathbf{A}$ in $\psi$ and again the elements of $\mathbf{A}$ are reordered such that $\psi$ remains invariant. The row vector *operating* on the left of $\mathbf{A}$ specifies the covariant indices and the column vector *operating* on the right of $\mathbf{A}$ specifies the contravariant indices in our convention; a covariant vector is a column vector and a contravariant vector is a row vector. Contraction of an upper and lower tensor index, which is a generalized trace operation,[6] corresponds to setting equal the corresponding vectors in $\psi$ and taking the sum of appropriate elements in $\mathbf{A}$.

Interchanging two indices in $A^{\beta_1\beta_2\cdots\beta_m}{}_{\alpha_1\alpha_2\ldots\alpha_k}$ and keeping $\psi$ invariant serves to remap either rows or columns of $\mathbf{A}$ in analogy with the elementary row and column operations.[7] For example, interchanging the $\alpha_i$ and $\alpha_j$ covariant indices of the tensor is equivalent to the following correspondences

$$(\mathbf{a}^T_{\alpha_1} \otimes \mathbf{a}^T_{\alpha_2} \otimes \cdots \otimes \mathbf{a}^T_{\alpha_i} \otimes \cdots \otimes \mathbf{a}^T_{\alpha_j} \otimes \cdots \otimes \mathbf{a}^T_{\alpha_k}) \, \mathbf{A} \, (\mathbf{b}_{\beta_1} \otimes \mathbf{b}_{\beta_2} \otimes \cdots \otimes \mathbf{b}_{\beta_m})$$

$$= (\mathbf{a}^T_{\alpha_1} \otimes \mathbf{a}^T_{\alpha_2} \otimes \cdots \otimes \mathbf{a}^T_{\alpha_j} \otimes \cdots \otimes \mathbf{a}^T_{\alpha_i} \otimes \cdots \otimes \mathbf{a}^T_{\alpha_k}) \, \mathbf{A}^{v_{ij}} \, (\mathbf{b}_{\beta_1} \otimes \mathbf{b}_{\beta_2} \otimes \cdots \otimes \mathbf{b}_{\beta_m})$$

$$\rightarrow \mathbf{A} = A^{\beta_1\beta_2\cdots\beta_m}{}_{\alpha_1\alpha_2\cdots\alpha_i\cdots\alpha_j\cdots\alpha_k}$$

$$\rightarrow \mathbf{A}^{v_{ij}} = A^{\beta_1\beta_2\cdots\beta_m}{}_{\alpha_1\alpha_2\cdots\alpha_j\cdots\alpha_i\cdots\alpha_k} \;\;,$$

where the notation $\mathbf{A}^{v_{ij}}$ designates a vertical row reordering interchanging those rows of $\mathbf{A}$ corresponding to those operated upon by the same indexed vectors on the left. Interchanging vectors on the right of $\mathbf{A}$ in $\psi$, i.e., $\beta_i$ and $\beta_j$, corresponds to a horizontal reordering of columns in $\mathbf{A}$ and this is designated as $\mathbf{A}^{h_{ij}}$. In each case the reordering of rows and/or columns in $\mathbf{A}$ is performed such that $\psi$ remains invariant. Furthermore, the vertical and horizontal transpose operations are commutative, and for two matrices $\mathbf{X}$ and $\mathbf{Y}$ and a row operation $\phi$, $\phi(\mathbf{XY}) = \phi(\mathbf{X})\mathbf{Y}$, and for a column operation $\mu$, $\mu(\mathbf{XY}) = \mathbf{X}\mu(\mathbf{Y})$.[7]

The above correspondence relations furnishes a rigorous connection between indicial tensor quantities and formal matrix mathematics. When considering some general tensor $A^{\beta_1\beta_2\cdots\beta_m}{}_{\alpha_1\alpha_2\ldots\alpha_k}$, from the point of view of doing numerical analyses, forming a array may be a straightforward matter of rearranging all of the components of the tensor (which may be treated as a km-dimensional matrix or array) into a two-dimensional field and then having the computer keep track of their disposition throughout the calculation. Our approach is significantly different in that we are interested in formulating problems, which hitherto have been treated entirely using tensor techniques and have appeared to be natural to tensor mathematics, fundamentally in formal matrix structures. In other words, we desire to cast the mathematical form of the physical problem as a *matrix equation* wherein the unknown quantity is a matrix; such equations may be nonlinear, algebraic, or differential. An important property of these equations is that the matrices comprising the factors in the terms do not commute so that the solution of these equations are extremely challenging. Matrix equations arise in many disciplines, including control theory, queuing theory,[8] and stability,[9] and even optics.[10]

The development of the field equations begins with the scalar quadratic form $ds^2 = d\mathbf{u}^T \mathbf{G} \, d\mathbf{u}$, where $\mathbf{G}$ is the symmetric $4\times 4$ metric tensor matrix with elements $g_{\mu\nu}$. Here $d\mathbf{u}^T = [du_1 \; du_2 \; du_3 \; du_4]$ are the coordinate differentials and $ds^2$ the line-element. Performing the standard variation along $ds$ yields the geodesic equation for the coordinate vector $\mathbf{u}$

$$\frac{d^2\mathbf{u}}{ds^2} + \Gamma \left( \frac{d\mathbf{u}}{ds} \otimes \frac{d\mathbf{u}}{ds} \right) = 0 \;\;,\;\; \Gamma = \mathbf{G}^{-1}\left[ [\frac{d\mathbf{G}}{du_1} \frac{d\mathbf{G}}{du_2} \frac{d\mathbf{G}}{du_3} \frac{d\mathbf{G}}{du_4}] - \frac{1}{2}\left(\frac{d\mathbf{G}}{d\mathbf{u}}\right)^T \right] \;\;, \tag{2}$$

where the vector derivative of $\mathbf{G}$ is defined such that the first column of $d\mathbf{G}/d\mathbf{u}$ is the derivative of the row-major ordered elements of $\mathbf{G}$ with respect to $u_1$, etc.[11] The matrix $\Gamma$ is a $4\times 16$ matrix analog to the Christoffel symbol, each $4\times 4$ block $d\mathbf{G}/du_i$ is the scalar derivative of $\mathbf{G}$ with respect to $u_i$. In the derivation of (2) the chain rule formula for derivatives of matrices with respect to a vector is used[12]

$$\frac{d(\mathbf{FGH})}{d\mathbf{u}} = (\mathbf{FG} \otimes \mathbf{I}_u)\frac{d\mathbf{H}}{d\mathbf{u}} + (\mathbf{F} \otimes \mathbf{H}^T)\frac{d\mathbf{G}}{d\mathbf{u}} + (\mathbf{I}_r \otimes \mathbf{H}^T\mathbf{G}^T)\frac{d\mathbf{F}}{d\mathbf{u}} \;\;, \tag{3}$$

where $\mathbf{F}$ is $r\times s$, $\mathbf{G}$ is $s\times t$ and $\mathbf{H}$ is $t\times u$, and $\mathbf{I}_j$ are $j\times j$ identity matrices. Here $\mathbf{G}$ is treated as a unique tensor independent of the above rules to enable the second derivative term to be isolated by allowing $\mathbf{G}^{-1}$ to exist.

The Einstein field equations are derived by taking the difference of the twice covariant differentiated scalar invariant $I = \mathbf{a}^T(d\mathbf{u}/ds)$ with its commuted derivatives for some vector $\mathbf{a}$ and contracting that tensor with respect to its one contravariant and a covariant index. In our formalism the Riemann curvature matrix is defined as

$$\frac{d}{ds_2}(\frac{d}{ds_1}I) - \frac{d}{ds_1}(\frac{d}{ds_2}I) = \left( \frac{d\mathbf{u}^T}{ds} \otimes \frac{d\mathbf{u}_1^T}{ds_1} \otimes \frac{d\mathbf{u}_2^T}{ds_2} \right) \mathbf{R} \, \mathbf{a} \;\;, \tag{4a}$$

where the subscripts on the ds and d**u** are used only to preserve ordering during the computation. All terms consisting of the vector derivatives of the vector **a** cancel out. The Riemann curvature matrix **R**, in direct analogy with the tensor form $R^\varepsilon_{\mu\nu\sigma}$, is

$$\mathbf{R} = (\Gamma^T \otimes \mathbf{I}_G)^{v_{23}} \Gamma^T - (\Gamma^T \otimes \mathbf{I}_G)\Gamma^T + \left(\frac{d\Gamma^T}{d\mathbf{u}}\right)^{BT v_{23}} - \left(\frac{d\Gamma^T}{d\mathbf{u}}\right)^{BT}, \qquad (4b)$$

where the vertical row reordering operation "$v_{23}$" is defined by the relation

$$\left(\frac{d\mathbf{u}^T}{ds} \otimes \frac{d\mathbf{u}_2^T}{ds_2} \otimes \frac{d\mathbf{u}_1^T}{ds_1}\right)(\Gamma^T \otimes \mathbf{I}_G) = \left(\frac{d\mathbf{u}^T}{ds} \otimes \frac{d\mathbf{u}_1^T}{ds_1} \otimes \frac{d\mathbf{u}_2^T}{ds_2}\right)(\Gamma^T \otimes \mathbf{I}_G)^{v_{23}}, \qquad (4c)$$

from which it is a straightforward matter to obtain $(\Gamma^T \otimes \mathbf{I}_G)^{v_{23}}$, and similarly the third term in (4b). For $\Gamma = [\Gamma_1\ \Gamma_2\ \Gamma_3\ \Gamma_4]$, which consists of 4×4 blocks $\Gamma_i$, $\Gamma^{BT} = [\Gamma_1^T\ \Gamma_2^T\ \Gamma_3^T\ \Gamma_4^T]$; we call $\Gamma^B$ the block-transpose of $\Gamma$ whereby the individual $\Gamma_i$ are not transposed; $\Gamma^T$ is the standard matrix transpose and $\mathbf{I}_G$ is a 4×4 identity matrix, having the same dimensions as the matrix **G**. Important formulae and identities used in this computation include Equation (3), and $\text{vec}(\mathbf{AXB}) = (\mathbf{B}^T \otimes \mathbf{A})\text{vec}(\mathbf{X})$, where $\text{vec}(\mathbf{X})$ is the column-major ordered form of any matrix $\mathbf{X}$,[4] and the Kronecker relationships $\mathbf{a}^T\mathbf{X}(\mathbf{M} \otimes \mathbf{b}) = \mathbf{b}^T\mathbf{X}^{BT}(\mathbf{M} \otimes \mathbf{a})$ and $(\mathbf{I}_G \otimes \mathbf{a}^T)\mathbf{X}^B = c(\mathbf{X}^B)(\mathbf{I}_G \otimes \mathbf{a})$, for conformal matrices **X** and **M**, and column vectors **a** and **b**, and $c(\mathbf{X}^B)$ is defined so that row i is $[\text{vec}(\mathbf{X}_i)]^T$, for a square block component $\mathbf{X}_i$ of **X**, where the blocks are arranged vertically within **X**. Note further that in Equation (2) that $c[(d\mathbf{G}/d\mathbf{u})]$ is the first term in $\Gamma$.

The matrix **R** has 64 rows and four columns corresponding to the three covariant indices and one contravariant index in $R^\varepsilon_{\mu\nu\sigma}$, as can be seen from comparing (4a) and (1). Contraction to obtain the Ricci tensor from $R^\varepsilon_{\mu\varepsilon\nu}$ entails -- for this choice of indices -- to taking the trace of each of the sixteen 4×4 blocks in **R**; these are the sixteen components of the vacuum field equations, and is ordered as a 16×1 matrix, which is designated as $\Theta$. $\Theta = 0$ are the vacuum field equations, and $\Theta$ is a covariant-indexed quantity as it is a 16-element column vector, i.e., its $\psi$-representation via (1) has vectors only on the left.

In applying a matrix formulation to the determination of the Riemann curvature tensor and hence leading to the Einstein vacuum field equations, one may ask why we even bother to discuss tensors if our intent is to remain fully within one domain. The reason that the $\psi$-representation arises, and quite naturally, is because of how the field equations are developed from derivatives of a scalar invariant associated with parallel transport of a vector along different geodesic trajectories in a curved space embodied in (4a). For $I = \mathbf{a}^T(d\mathbf{u}/ds)$ the following expressions are readily derived

$$\frac{dI}{ds} = \frac{d\mathbf{u}^T}{ds}\left\{-\Gamma^{BT}(\mathbf{I}_G \otimes \mathbf{a}) + \frac{d\mathbf{a}}{d\mathbf{u}}\right\}\frac{d\mathbf{u}}{ds}$$

$$= \left\{ -\mathbf{a}^T \Gamma + \text{vec}(\frac{d\mathbf{a}}{d\mathbf{u}})^T \right\} \left( \frac{d\mathbf{u}}{ds} \otimes \frac{d\mathbf{u}}{ds} \right)$$

Each expression is equally valid and signifies a second-rank tensor of the mixed form $A_\mu{}^\nu$ in the first case and the covariant tensor $A_{\mu\nu}$ in the second case. The transpose of each expression is also a scalar and corresponds to the mixed tensor $A^\mu{}_\nu$, and the contravariant tensor $A^{\mu\nu}$, respectively. Interchanging the imaginary indexed order of the coordinate derivatives outside the brackets is in keeping with the above rules of index switching and hence the two indices in this second rank tensor. It is important that only one representation be adopted throughout a derivation, such as that leading to (4b), and the proper implementation of a contraction yields the system of equations which interest us. Contraction on the $\psi$-representation is a generalized trace operation with respect to the chosen vector on the left and on the right of $\mathbf{A}$, whereby the contracted matrix has 1/4th the number of rows and columns. That is, contraction may be performed on a rectangular matrix such as the Riemann curvature matrix $\mathbf{R}$ as it appears in (4a). For the first equation above, corresponding to $A_\mu{}^\nu$, contraction is simply the trace of the bracketed quantity. All of this goes to show the motivations and power and flexibility of tensor mechanics, though the $\psi$-representation appears to furnish a bridge between formal matrix mathematics and tensor mechanics, which to our knowledge has not hitherto been clearly delineated in the physics literature.

    Unfortunately the highly organized and rigid rules of matrix mechanics has a number of drawbacks. The first is that the derivation leading to (4a) makes use of non-intuitive derivative expressions such as the chain rule in (3). The second is that in order to keep all terms consistent in regards to the $\psi$-representation it is necessary to use complex Kronecker matrix identities and to adopt usage of a number of unconventional matrix representations, i.e., the vertical and horizontal reordering operations. The third is that many of the familiar identities with the Riemann tensor and its derivatives (the Bianchi identities) as well as the definition of the Weyl tensor are not obvious or natural to the matrix formalism. The major advantage resides in the compactness of the resulting equations for numerical work; possibly new inferences about the structure of the field equations can be made from more detailed examination of our expressions utilizing the substantial theory developed in linear and multilinear algebra. One such, perhaps already quite obvious, inference is that the definition of $\Gamma$ and the contracted Riemann expression form a system of matrix equations. The first is a linear differential equation in $\mathbf{G}$ and the latter a Riccati-type equation in $\Gamma$. Substantial work on the existence and determination of solutions of matrix Riccati equations exists,[13] and as such our representation of the field equations may enable new mathematical techniques to be brought to bear on the struggle to crack the field equations for various physical problems. The matrix expression are also much easier to handle with regard to implementing numerical algorithms using many standard software packages (IDL, Matlab, and Mathematica, for example) even with our unconventional reordering operations. Conceivably the

matrix equation formulation may be useful for developing algorithms to tackle the equivalence problem in general relativity involving finding practical algorithms or analytic tests to show when two different metrics are identical even though they are expressed different coordinate systems.[14]

The author wishes to thank Eric Le Bigot for his very helpful discussions on this topic.

# References

1. Landau, L.D., and E.M. Lifshitz, The Classical Theory of Fields (Pergamon Press, Oxford, 1975, 4th Revised English Edition), p. 263.
2. Flanders, H., *Differential Forms with Applications to the Physical Sciences* (Dover, New York, 1989).
3. Newman, E.T., and R. Penrose, J. Math. Phys. **3**, 566-79 (1962). See also Ernst, F.J., J. Math. Phys. **15**, 1409-12 (1974).
4. Bell, E.T., *The Development of Mathematics* (McGraw-Hill, New York, 1940), p. 188.
5. Horn, R.A., and C.R. Johnson, *Topics in Matrix Analysis* (Cambridge Univ. Press, 1991).
6. Bishop, R.L., and S.L. Goldberg, *Tensor Analysis on Manifolds* (Dover, New York, 1980).
7. Mirsky, L., *An Introduction to Linear Algebra* (Dover, New York, 1990), p. 170.
8. Neuts, M.F., *Matrix-Geometric Solutions in Stochastic Models* (Dover, New York, 1981).
9. Gajic, Z., and M. Qureshi, *Lyapunov Matrix Equation in System Stability and Control* (Academic Press, San Diego, 1995).
10. Wild, W.J., Optics Letters **21**, 1433 (1996).
11. G. S. Rogers, *Matrix Derivatives* (Marcel Dekker, Inc., New York, 1980). We adapt the row-major ordering for matrix derivatives, though a column-major ordering can be just as easily adapted, so long as consistency is retained throughout the derivation.
12. Marlow, W.H., *Mathematics for Operations Research* (Dover, New York, 1978).
13. Bittanti, S., A.J. Laub, and J.C. Willems, Eds, *The Riccati Equation* (Springer-Verlag, Berlin 1991).
14. Karlhede, A., Gen Rel & Grav. **12**, 693 (1980).